\documentclass[twocolumn,aps,showpacs,superscriptaddress,prl]{revtex4}

\usepackage{amsmath,amssymb,amsmath}
\usepackage{graphicx}
\usepackage{dcolumn}
\usepackage{bm}

\begin{document}
\title{Comparative study of the rovibrational properties of heteronuclear alkali
  dimers in electric fields}

\author{Rosario Gonz\'alez-F\'erez}
\email{rogonzal@ugr.es}
\affiliation{Instituto 'Carlos I' de F\'{\i}sica Te\'orica y
Computacional and Departamento de F\'{\i}sica At\'omica Molecular y
Nuclear, Universidad de Granada, E-18071 Granada, Spain}

\author{Michael Mayle}
\affiliation{Theoretische Chemie, Physikalisch--Chemisches
Institut, Universit\"at Heidelberg,
Im Neuenheimer Feld 229, D-69120 Heidelberg, Germany}

\author{Pablo S\'anchez-Moreno}
\affiliation{Instituto 'Carlos I' de F\'{\i}sica Te\'orica y
Computacional and Departamento de F\'{\i}sica At\'omica Molecular y
Nuclear, Universidad de Granada, E-18071 Granada, Spain}

\author{Peter Schmelcher}
\affiliation{Theoretische Chemie, Physikalisch--Chemisches
Institut, Universit\"at Heidelberg,
Im Neuenheimer Feld 229, D-69120 Heidelberg, Germany}
\affiliation{Physikalisches Institut, Universit\"at Heidelberg,
Philosophenweg 12, D-69120 Heidelberg, Germany}

\date{\today}

\begin{abstract}
A comparative study of the effect of a static homogeneous electric
field on the rovibrational spectra of several polar dimers in their
$\textrm{X}^1\Sigma^+$ electronic ground state is performed.
Focusing upon the rotational ground state within each vibrational band,
results for energies and various expectation values are presented.
For moderate field strengths the electric field-induced energy shifts,
orientation, alignment, and 
angular motion hybridization are analyzed up to high vibrational excitations
close to the dissociation threshold.
\end{abstract}


\maketitle

\section{Introduction}
The availability of cold and ultracold dimers provides a unique
opportunity to investigate fundamental quantum processes \cite{speciss2004}. 
The efforts of many experimental groups are focused on the production of
heteronuclear alkali dimers, as they constitute a prototype of an ultracold
quantum gas with long-range dipole-dipole interactions. In this context, the
photoassociation of LiCs, NaCs, KRb, and RbCs has been reported recently
\cite{kraft06,haimberger04,sage:203001,wang04}, while Feshbach resonances have
been observed for LiNa, LiK, LiRb, and KRb
\cite{stann05,deh08,wille08,hodby05,ospelkaus}. 
The long-range anisotropic dipole-dipole interaction between these polar
molecules gives rise to an intriguing many body physics with new interesting
physical phenomena. 
In addition, these systems allow for a wide range of applications, such
as state resolved chemical reactions  
\cite{bodo02} or quantum computing \cite{demille02,yelin06}. 

Particularly interesting is the study of the influence of external fields on
these molecular systems. 
External fields play a key role in trapping and cooling
processes as well as in the manipulation and control of the
dipole-dipole interaction, chemical reactions or collisions
\cite{krems06}.
As an example, the application of an electric field has been predicted to
significantly affect the photoassociation process of ultracold  
polar molecules as well as the subsequent radiative decay cascade \cite{gonzalez07,gonzalez07_2}.
On the other hand, the internal structure of molecules is significantly
affected by an electric field.   
The strong field regime is characterized by the appearance of pendular
states: The molecule is oriented along the field axis and
can be described as a coherent superposition of field-free rotational states.
Recently, the authors have performed studies of the rovibrational 
spectra of CO and LiCs in their electronic ground states exposed to an electric
field: The orientation and hybridization of the 
angular motion, the Stark shift, and the radiative decay
properties have been investigated for a wide range of field strengths and rovibrational 
levels \cite{gonzalez04,gonzalez06,mayle06}.   
In the present work we extend our previous studies by performing a
comparative analysis of several polar alkali dimers
which are of immediate experimental interest. Specifically, we 
analyze the influence of a static electric field on the $\textrm{X}^1\Sigma^+$
electronic ground state of $^6$Li$^{23}$Na, $^6$Li$^{40}$K, $^6$Li$^{87}$Rb,
$^7$Li$^{133}$Cs, $^{23}$Na$^{39}$K, $^{23}$Na$^{133}$Cs,
$^{39}$K$^{85}$Rb, and $^{85}$Rb$^{133}$Cs.   

\section{Rovibrational Hamiltonian}
In the following we consider the field regime where perturbation theory holds
for the description of the molecular electronic structure. 
The Born-Oppenheimer
approach then provides the following Hamiltonian for the nuclear motion
of a polar dimer 
in its electronic ground state exposed to a static electric field
\begin{equation}
\label{eq:rotvib_hamiltonian}
H= T_R
+\frac{\hbar^2\mathbf{J}^2(\theta,\phi)}{2\mu R^2} +
\varepsilon(R)-FD(R)\cos\theta 
\end{equation}
where the molecule fixed frame with the origin at the center of
mass of the nuclei has been employed, ($R,\theta,\phi$) being the
internuclear distance and the Euler angles\footnote{The third Euler angle is not required since 
rotations around the molecular axis are not taken into account.}.
$T_R$, $\mu$,
$\hbar\mathbf{J}(\theta,\phi)$, $\varepsilon(R)$, and $D(R)$ are the
vibrational kinetic energy, the reduced mass of the nuclei, the rotational
angular momentum, the field-free electronic potential energy curve (PEC), and
the electronic dipole moment function (EDMF), respectively. The electric field
is taken parallel to the laboratory frame $z$-axis with strength $F$. 
Since we restrict ourselves on a non-relativistic treatment,
couplings to electronic states of different spin symmetry do not
arise.

In the framework of the effective rotor approach \cite{gonzalez04}, the
rovibrational Hamiltonian (\ref{eq:rotvib_hamiltonian}) reduces to   
\begin{equation}
\label{hamil_era}
 H_\nu^{ERA}=\frac{\hbar^2}{2\mu}\langle R^{-2}\rangle_\nu^{0}
\mathbf{J}^2 -F\langle D(R)\rangle_\nu^{0}\cos\theta +E_\nu^{0}
\end{equation}
with $\langle R^{-2}\rangle_\nu^{0}=\langle\psi_\nu^{0}|R^{-2}|
\psi_\nu^{0}\rangle$,
$\langle D(R) \rangle_\nu^{0} = \langle\psi_\nu^{0}|D(R)|\psi_\nu^{0}\rangle$. 
$\psi_\nu^{0}$ and
$E_\nu^{0}$ are the field-free vibrational wave function and energy 
for a given vibrational state $\nu$ with $J=0$, respectively.
This approximation, which is valid for large parts of the 
rovibrational spectrum, relies on two assumptions:
i) the rotational and vibrational energy scales should differ by several
orders of magnitude and can therefore be separated adiabatically, and
ii) the field influence on the vibrational motion is
very small and can consequently be treated by perturbation theory.
Besides reducing (\ref{eq:rotvib_hamiltonian}) to an one-dimensional problem, 
this approach allows us to compare the field influence on several
vibrational levels, and even between different molecular species,  
by means of the parameter
\begin{equation}
  \label{eq:omega_nu}
  \omega_\nu= \frac{2\mu \langle D(R) \rangle_\nu^{0}}
{\hbar^2\langle R^{-2} \rangle_\nu^{0}},  
\end{equation}
which derives from rescaling the Hamiltonian (\ref{hamil_era}) using
the rotational constant of a given vibrational state, 
$B_\nu=\hbar^2 \langle R^{-2} \rangle_\nu^{0}/2\mu$.
If $\omega_\nu\cdot F$ acquires the same order of magnitude as the 
rescaled rotational kinetic energy, $\mathbf J^2$, the rotational
dynamics is severely altered by the interaction with the electric field, 
while $\omega_\nu^{-1}$ provides an estimation of the field strength needed to 
show a significant effect.

We remark that in the presence of the electric field, only the magnetic
quantum number is retained, 
whereas the vibrational and rotational ones are no more conserved.
However, for reasons of addressability, we will label the
electrically dressed states by means of their field-free vibrational,
rotational, and magnetic quantum numbers $(\nu,J,M)$. 
Furthermore, in this letter we focus on rotationally non-excited states
with corresponding field-free quantum numbers $J=M=0$.
The computational method used to solve the full rovibrational Schr\"odinger
equation is described in ref.~\cite{gonzalez04}. 

\section{Electronic potential curves and dipole moment functions}
The PECs of the singlet electronic ground states of the 
alkali dimers employed here are derived from
either experimental or theoretical studies. For the PECs of
LiNa, LiK, LiCs, NaK, NaCs, KRb, and RbCs we use very accurate
experimental data from spectroscopic analyses
\cite{martin01,fellows91,staanum06,russier00,Docenko06,pashov07,fellows99}.
The LiRb PEC is taken from theoretical studies based on a semiempirical
method \cite{korek00}.
When describing vibrationally highly excited
states, special attention has to be payed to the long-range part of the PECs:
For LiNa, LiCs, NaCs, KRb, and RbCs, the PECs include van der Waals terms, 
$-\sum_{n=6,8,10}C_n/R^{n}$, and an exchange energy term,  
$-AR^\gamma e^{-\beta R}$, see 
refs.~\cite{martin01,fellows91,staanum06,Docenko06,pashov07,fellows99},
with $\gamma=0$ for LiNa \cite{fellows91}. The long-range part of the NaK PEC
additionally includes damping functions $\xi_n(R)$ in the van der Waals
expansion, 
$-\sum_{n=6,8,10}\xi_n(R)C_n/R^{n}$, and a different expression 
for the exchange energy, $-A e^{-\alpha R-\beta R^2}$ \cite{russier00}. For
the LiRb dimer only the first van der Waals term, $-C_6/R^{6}$, is 
considered \cite{Derevianko01}.

\begin{figure}
\includegraphics[scale=.8]{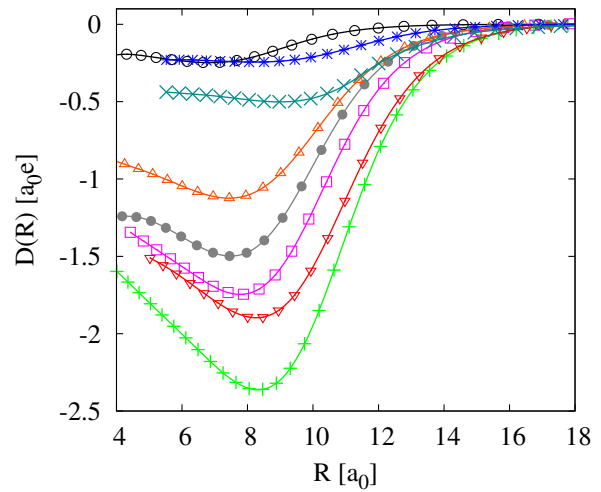}
\caption{Color online. Electric dipole moment functions of 
the $\textrm{X}\,^1\Sigma^+$ electronic  ground states of LiNa ($\circ$), 
  LiK ($\bullet$),  LiRb ($\Box$),  LiCs ($+$), NaK ($\triangle$),  
  NaCs($\triangledown$), KRb ($\ast$), and RbCs ($\times$).} 
\label{fig:edmf}
\end{figure}
 
The EDMFs employed for the LiNa, LiK, LiCs, NaK, NaCs, and KRb systems
are computed by the same semiempirical technique as the 
LiRb PEC \cite{aymar05}. The RbCs EDMF is taken from an ab-initio relativistic
valence-bound calculation \cite{koto05}.
Since for the electronic ground state of the polar alkali dimers the 
long-range behaviour of the EDMF is given by $D_7/R^{7}$ 
\cite{byers70}, this asymptotic function has been fitted to the 
theoretical data at large internuclear distances.
We note that, when necessary for computational reasons, the short-range behaviour
of the EDMFs is linearly extrapolated. The resulting EDMFs are plotted in
fig.~\ref{fig:edmf} as a function of the internuclear distance.
All EDMFs are negative and exhibit a
broad minimum which is shifted with respect to the minimum of the corresponding
PEC, {\it e.g.}, by $0.51$ and $1.39\, a_0$ for the KRb and LiCs dimers,
respectively. 
The EDMFs of LiRb, NaK, NaCs, and KRb become positive for $R\gtrsim 17.3\,a_0$
albeit approaching zero thereafter;
this change of sign will be reflected in the response of highly 
excited rovibrational levels to the field.
However, this has to be taken with a grain of salt since ab-initio electronic
structure calculations of molecules typically employ basis sets of
exponentially localized functions, which do not easily catch the long-range
behaviour of the atom-atom interaction.
The LiCs EDMF has the largest absolute value with a maximum of 
$|D(R)|=2.36\,a_0e$ at $R=8.3\,a_0$. 
The EDMFs of LiNa, KRb, and RbCs show only a weak dependence
on $R$ for a wide range of internuclear distances and overall comparatively
small values for the dipole moment.

\begin{figure}
\includegraphics[scale=0.8]{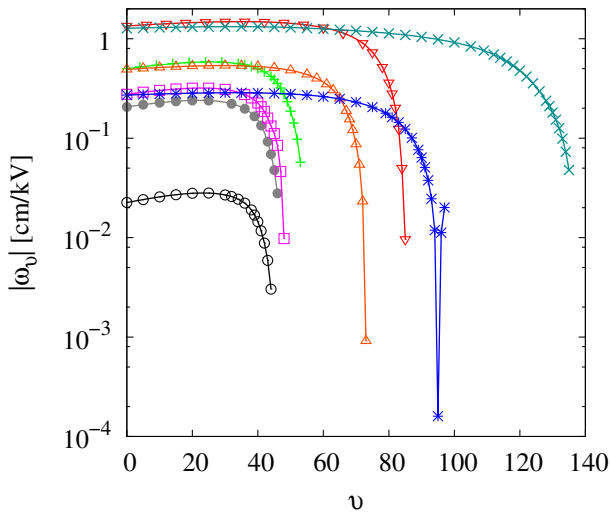}
\caption{Color online. The parameter $|\omega_\nu|$ versus the field-free
vibrational quantum  number $\nu$ for the same dimers as in  fig.~\ref{fig:edmf}.} 
\label{fig:omega_nu} 
\end{figure}

\section{Results}
Let us start by analyzing the behaviour of the parameter $\omega_\nu$ as a
function of the vibrational quantum number. In fig.~\ref{fig:omega_nu}, the
modulus of $\omega_\nu$ is presented for the $(\nu,0,0)$ levels of all dimers
on a logarithmic scale. The maximum vibrational quantum numbers 
considered  are $\nu=44$ (LiK), 46 (LiNa), 48 (LiRb), 53 (LiCs), 73 (NaK),
85 (NaCs), 97 (KRb), and 135 (RbCs);
the very last vibrational bands have not been included. For all dimers, 
$|\omega_\nu|$ shows a qualitatively similar but quantitatively different
behaviour as a function of $\nu$:
After a weakly pronounced maximum, $|\omega_\nu|$ decreases rapidly for high vibrational
excitations. In particular, NaK, KRb, and RbCs show a plateau-like behaviour
indicating a quantitatively similar field impact in many vibrational bands. 
According to their $|\omega_\nu|$ values, the considered molecules 
can be classified into four groups whose response to the  
field will be of comparable magnitude for a large part of their spectra.
Possessing the largest $|\omega_\nu|$ values, the first group is formed by the
NaCs and RbCs dimers and their $\nu\lesssim65$ states, followed by the
$\nu\lesssim 40$ levels of LiCs and NaK. 
Despite the large LiCs EDMF, the corresponding $|\omega_\nu|$ 
is only approximately half of the $|\omega_\nu|$ belonging to NaCs and RbCs:
The large reduced masses of the latter compensate their smaller dipole moments.
The third set contains LiK, LiRb, and KRb and their lowest $40$
vibrational bands. Finally, LiNa will be least affected by the field
which is due to both small $D(R)$ and $\mu$.
We remark that the sign of $\omega_\nu$ depends crucially on 
the corresponding dipole moment: While being generally negative, 
the change of sign of the
NaK, NaCs, and KRb EDMFs at long distances causes $\omega_\nu>0$ for
their $\nu\ge 73$, $85$ and $95$ levels, respectively.
In particular, the minimum observed for the $\nu=95$ level of 
KRb can be attributed to this change of sign.
For all molecules considered, the slope of $|\omega_\nu|$ becomes very steep
as $\nu$ 
approaches the dissociation threshold. Consequently $|\omega_\nu|$ drastically
varies  from its maximum to its minimum; for example we get
for LiNa $\omega_{26}=-2.8\times 10^{-2}$ versus $\omega_{46}=-3.0\times
10^{-3}$ cm/kV, and 
for NaCs $\omega_{35}= -1.49$ versus $\omega_{85}=-9.6\times 10^{-3}$ cm/kV.
Hence, for high-lying vibrational excitations the field strength 
should be enhanced at least one order of magnitude
to obtain similar field-dressed states as on the lower part of
the rovibrational spectrum.    

In order to quantify the spectral changes upon the application of the electric
field, we consider the ratio of the Stark shift to the rotational 
spacing in the presence of the field
\begin{equation}
  \label{eq:kappa}
  \kappa_\nu=\frac{E_{\nu 0}(0) -E_{\nu 0}(F) }{E_{\nu 1}(F) -E_{\nu 0}(F) }
\end{equation}
where $E_{\nu 0}(F)$ and $E_{\nu 1}(F)$ are the energies of the $(\nu,0,0)$
and $(\nu,1,0)$ levels at field strength $F$;
for high-field seekers we get $\kappa_\nu>0$.
If $\kappa_\nu \ll 1$, the field-dressed spectrum shows many similarities
to the field-free one. In particular, this criterion is fulfilled for
$\omega_\nu F\ll 1$: In second order perturbation theory the $(\nu,0,0)$
and $(\nu,1,0)$ states are shifted by $-\omega_\nu^2F^2B_\nu/6$ and
$\omega_\nu^2F^2B_\nu/6$ \cite{meyenn}, respectively, compared to 
the field-free rotational spacing of $2B_\nu$.

\begin{figure}
\includegraphics[scale=0.8]{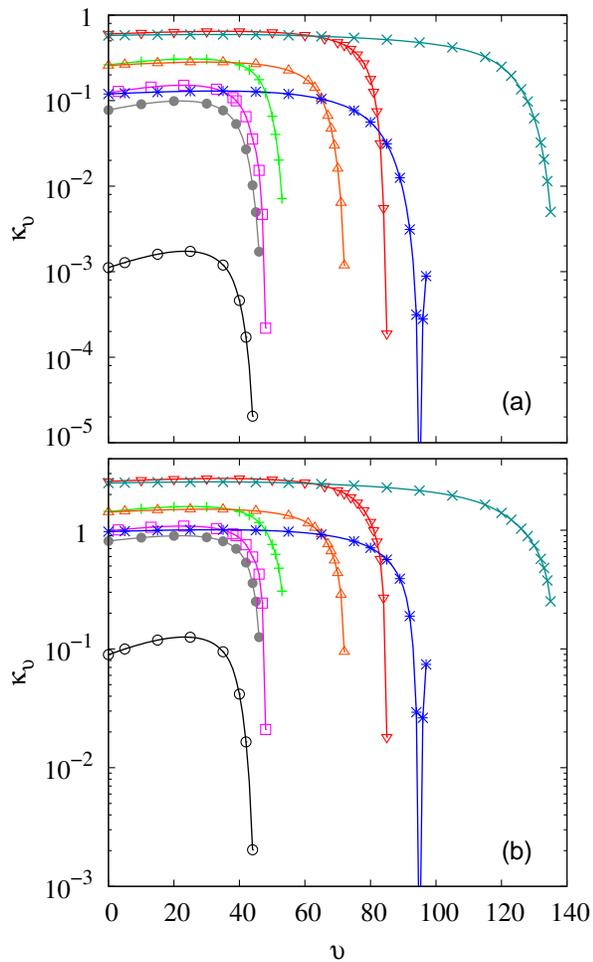}
\caption{Color online. The relative energy shift $\kappa_\nu$ as a function of
  the   field-free vibrational quantum number for the $(\nu,0,0)$ levels and
same molecules as  in   
  fig.~\ref{fig:edmf}, for (a) $F=5.14$ kV/cm and (b) $51.4$ kV/cm.} 
\label{fig:kappa} 
\end{figure}

In figs.~\ref{fig:kappa}(a) and (b) the parameter $\kappa_\nu$ is presented
as a function of the vibrational quantum number for $F=5.14$ and $51.4$ kV/cm,
respectively;
the selected field strengths are chosen such that 
the experimentally accessible range within the strong field regime is covered.
Besides showing high-field seeking character, the behaviour of $\kappa_\nu$
has many analogies among the considered molecules. 
Initially, it has a weak dependence on the vibrational excitation, thereby
smoothly increasing as $\nu$ is  
enhanced and passing a shallow maximum (due to the semilogarithmic
scale, this increasing trend appears as a plateau).
For higher vibrational excitations,
$\kappa_\nu$ rapidly decreases as the levels get closer to the
dissociation limit: This reduction can be up to one order of magnitude when
$\nu$ is enhanced by merely one quantum.  The above classification of the
considered molecules is equally observed  for $\kappa_\nu$. The lowest values
appear for LiNa, $\kappa_\nu\le 2\times 10^{-3}$ and $0.2$, 
for $F=5.14$ and $51.4$ kV/cm, respectively, {\it i.e.}, the impact of the field
on the spectrum is negligible for the smaller of these two field strengths.
In contrast, NaCs and RbCs are most affected by the field $F=5.14$ kV/cm, with 
$\kappa_\nu>0.5$ for $\nu\le68$ and $89$, respectively. LiCs and NaK possess
values $\kappa_\nu>0.2$ for $\nu\le 44$ and $58$, respectively.
For LiRb and KRb we found $\kappa_\nu>0.1$ for the $\nu\le38$ and $67$
states, whereas a smaller value is achieved for LiK
with $\kappa_\nu>0.08$ for $\nu \le34$.  An increase of the field strength by one
order of magnitude ($F=51.4$ kV/cm) produces a significant variation of $\kappa_\nu$, which
depends on the system and especially on the states under consideration.
Indeed, we observe $\kappa_\nu> 1$ for most levels of LiCs, NaK, NaCs, and
RbCs. The levels $\nu \le 36, 41$,  and $75$ of LiK, LiRb, and KRb satisfy
$\kappa_\nu >0.8$. 
In order to provide a reasonable scaling, 
the minimum of $\kappa_\nu$ for KRb at $\nu=95$ 
has not been included in fig.~\ref{fig:kappa}.

\begin{figure}
\includegraphics[scale=0.8]{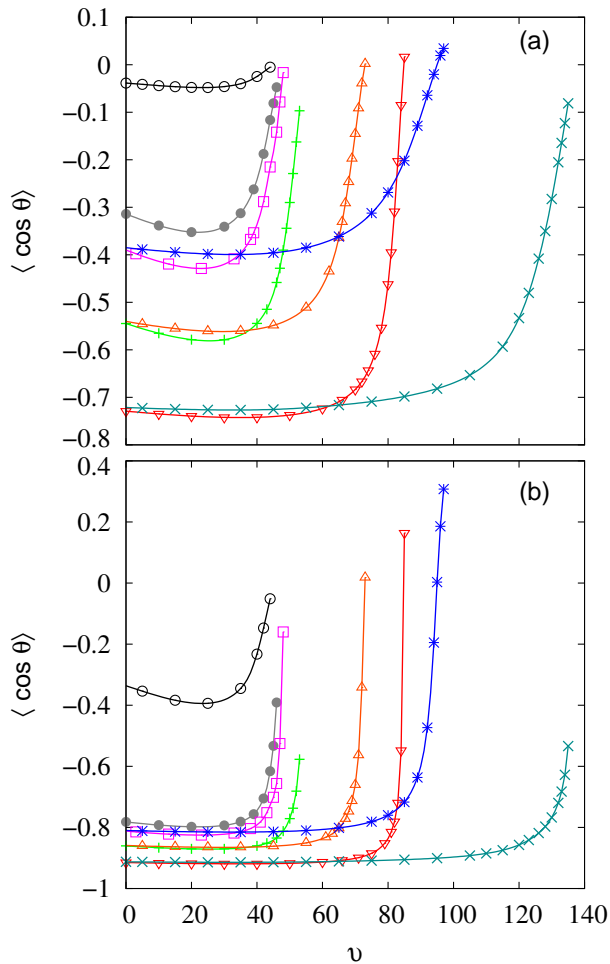}
\caption{Color online. The same as fig.~\ref{fig:kappa} but for the 
expectation value $\langle\cos\theta\rangle$.}    
\label{fig:cos}
\end{figure}

\begin{figure}
\includegraphics[scale=0.8]{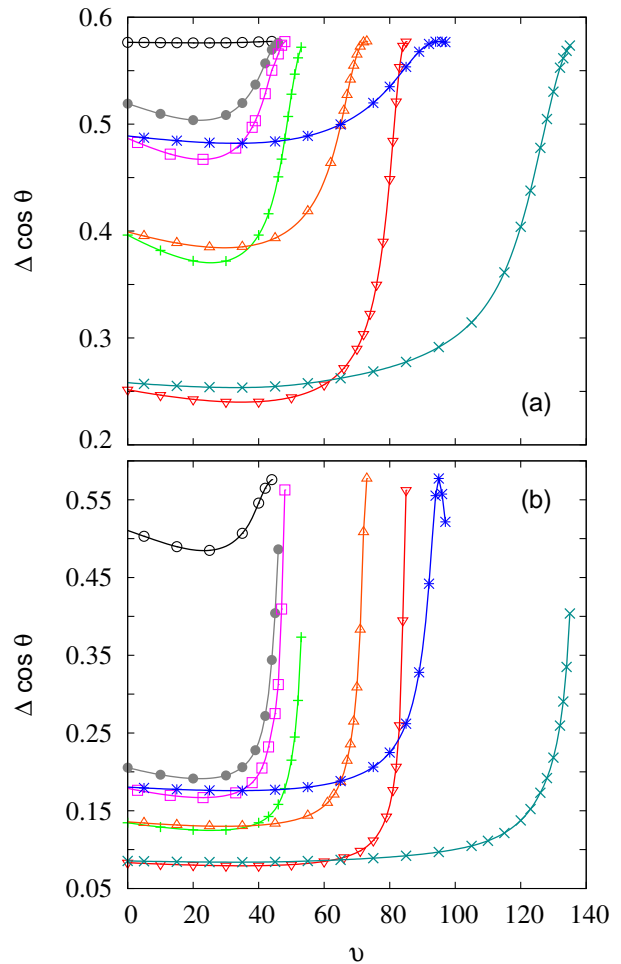}
\caption{Color online. The same as fig.~\ref{fig:kappa} but for 
$\Delta \cos\theta$.} 
\label{fig:delta} 
\end{figure}

The orientation and alignment of the pendular states are
characterized by the expectation value $\langle\cos\theta\rangle$ and
$\Delta\cos\theta=\sqrt{\langle\cos^2\theta\rangle-\langle\cos\theta\rangle^2}$, 
respectively. The closer $\Delta\cos\theta$ is to zero, the stronger is the
alignment and the closer the modulus of $\langle\cos\theta\rangle$ is
to one, the stronger is the orientation of the state along the 
field. The expectation value $\langle\cos\theta\rangle$ of these levels is
presented in figs.~\ref{fig:cos}(a) and (b) 
for $F=5.14$ and $51.4$ kV/cm, respectively, as a function of the
field-free vibrational quantum number. The corresponding results for 
$\Delta\cos\theta$ are shown in figs.~\ref{fig:delta}(a) and (b).
The sign of the effective electric dipole moment, 
$\langle D(R)\rangle_\nu^{0}$, determines whether the orientation will be
parallel or antiparallel to the field direction, 
{\it i.e.}, $\langle\cos\theta\rangle>0$ or $\langle\cos\theta\rangle<0$,  
respectively.  

For the moment let us focus on the description of the results for 
$F=5.14$ kV/cm, see figs.~\ref{fig:cos}(a) and \ref{fig:delta}(a).
As expected from the behaviour of the parameter $\omega_\nu$,
the orientation smoothly increases as $\nu$ is enhanced, reaching a broad
maximum and rapidly decreasing thereafter as the state gets close to the
dissociation threshold. Since the initial increasing trend is not very
pronounced for NaK, NaCs, KRb, and RbCs, $\langle\cos\theta\rangle$
exhibits a plateau as a function of $\nu$ followed by 
a rapid decrease of the orientation. 
The strongest orientation is achieved for the NaCs and RbCs dimers with 
$\langle\cos\theta\rangle\le-0.7071$, 
for the
$\nu\le 65$ and $76$ levels, respectively. 
The states $\nu\le 44$ and $56$ of LiCs and NaK, respectively,
present a significant orientation as well: 
$\langle\cos\theta\rangle\lesssim-0.5$. 
For LiK, LiRb, and KRb the influence of the field is not very
pronounced and the extremal values are
$\langle\cos\theta\rangle=-0.35$, $-0.43$, and $-0.40$ for the 
$\nu=22$, $23$, and $33$ levels, respectively. 
For LiNa as well as for highly vibrationally excited levels of the other
molecules, 
the effect of an electric field of strength $F=5.14$ kV/cm is negligible.
Analogous conclusions are derived from the analysis of
$\Delta\cos\theta$, whose behaviour as a function of $\nu$ is   
very similar to the one previously described  for $\langle\cos\theta\rangle$. 
The variation of $\Delta\cos\theta$ is very smooth as $\nu$ is augmented, and
major changes are observed only for high-lying excitations.   
The NaCs and RbCs dimers show the strongest alignment, with 
$\Delta\cos\theta < 0.3$ for the $\nu\le 72$ and $99$ levels, respectively.
The other dimers present relative large values with
$\Delta\cos\theta> 0.3703$ which is the lowest one achieved for LiCs
at $\nu=25$, indicating a modest alignment.  
In particular, $\Delta\cos\theta$ is very close to its field-free value,
{\it i.e.}, $\Delta\cos\theta\approx0.577$, for the LiNa dimer and for all
high-lying  vibrational levels. 

If the field strength is increased to $F=51.4$ kV/cm, both the orientation
and alignment are more pronounced, see figs.~\ref{fig:cos}(b) and
\ref{fig:delta}(b).
Except for the LiNa molecule, the strong field regime
with the appearance of pendular states is already reached for
most of the levels considered.
We observe $\langle\cos\theta\rangle\lesssim -0.8$ for the states with
$\nu\le 38, 48, 65, 81$, and $108$ of the LiRb, LiCs, NaK, NaCs, and RbCs
dimers, respectively. Even more, $\langle\cos\theta\rangle\lesssim -0.9$ for
the $\nu\le 70$ and $96$ states of NaCs and RbCs, respectively.
The LiNa molecule still shows a modest orientation, with a minimal value 
$\langle\cos\theta\rangle=-0.394$ for the $\nu=23$ state: only if $F$ is
further increased by one order of magnitude a significant orientation,
{\it i.e.}, $\langle\cos\theta\rangle\lesssim-0.8$, is attained for this
molecule.  
For $F=51.4$ kV/cm, the strong field regime is also manifest in 
$\Delta\cos\theta$ where many levels present a pronounced alignment with
$\Delta\cos\theta< 0.3$. The smallest values, 
$\Delta\cos\theta\lesssim 0.1$, are obtained for the $\nu\le 71$ and $100$
states of NaCs and RbCs, respectively. In contrast, for the LiNa
dimer we find that $\Delta\cos\theta\gtrsim 0.485$ for any $\nu$.  
Finally, let us remark that $\langle\cos\theta\rangle$ is positive for the
states with $\nu\ge 73$, $85$, and $95$ of the NaK, NaCs, and KRb dimers,
respectively, while being characterized by a weak orientation and alignment
at both field strengths. In particular, for KRb $|\langle\cos\theta\rangle|$ 
($\Delta\cos\theta$) achieves a minimum (maximum) for the $\nu=95$ level,
{\it i.e.}, for the same state as $|\omega_\nu|$ shows a minimum, cf.\ 
fig.~\ref{fig:omega_nu}.

\begin{figure} 
\includegraphics[scale=0.8]{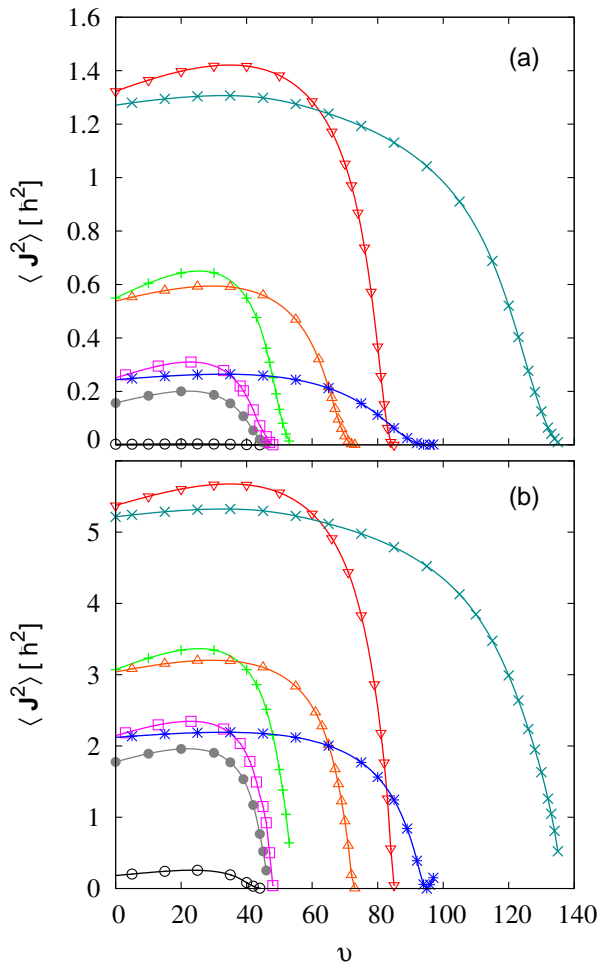}
\caption{Color online. The same as fig.~\ref{fig:kappa} but for the 
expectation value $\langle\mathbf{J}^2\rangle$.} 
\label{fig:j2} 
\end{figure}

The expectation value $\langle\mathbf{J}^2\rangle$ provides an estimate of
the field-induced hybridization of the angular motion. 
Since we only consider the rotational ground state in this letter,
one yields $\langle\mathbf{J}^2\rangle=0$ for $F=0$.
Figures \ref{fig:j2}(a) and (b) illustrate the evolution of
$\langle\mathbf{J}^2\rangle$
with increasing degree of vibrational excitation 
for $F=5.14$ and $51.4$ kV/cm, respectively.  
For both field strengths investigated, $\langle\mathbf{J}^2\rangle$ presents a
similar behaviour for all molecules:
It increases as the vibrational excitation is
enhanced, passes through a broad maximum, and decreases, approaching zero,
thereafter.
The lower parts of the NaK, KRb, and RbCs spectra 
show again a plateau-like behaviour of $\langle\mathbf{J}^2\rangle$.
For $F=5.14$ kV/cm, the rotational dynamics has mostly 
$s$-wave character, {\it e.g.}, $47\%$  for the $\nu=34$ level of
NaCs, $73\%$ for the $\nu=29$ state of NaK, and $87\%$ for the $\nu=33$ 
level of KRb.
The hybridization of the angular motion is negligible for LiNa, {\it i.e.}, 
$\langle\mathbf{J}^2\rangle<3.5\times 10^{-3}\, \hbar^2$  for
any $\nu$, and $99.8\%$ of the dynamics is controlled by the $J=0$ wave.  
In contrast, the pendular character of the NaCs and RbCs dimers is already
significant with $\langle\mathbf{J}^2\rangle>1.0 \, \hbar^2$ for the $\nu\le 71$
and $98$  
levels, respectively. 
The $\nu\le 42$ and $52$ states of LiCs and NaK, respectively, satisfy
$\langle\mathbf{J}^2\rangle>0.5 \, \hbar^2$; analogously we observe for LiRb
and KRb  
$\langle\mathbf{J}^2\rangle>0.2 \, \hbar^2$ for $\nu\le 39$ and $67$.
For the LiK states we find $\langle\mathbf{J}^2\rangle>0.15 \, \hbar^2$  for
$\nu\le 35$. 

Because of the high-field seeking character of these levels, 
$\langle\mathbf{J}^2\rangle$ is drastically increased for $F=51.4$ kV/cm, see 
fig.~\ref{fig:j2}(b).  
Except for LiNa, strong field effects characterize this
expectation value for a wide range of levels. Indeed,
$\langle\mathbf{J}^2\rangle>5.0 \, \hbar^2$ 
for the $\nu \le 64 $ and $73$ states of NaCs and RbCs, 
$\langle\mathbf{J}^2\rangle>3.0 \, \hbar^2$ for those with $\nu \le 50 $ and
$41$ of the NaK and LiCs dimers, and 
$\langle\mathbf{J}^2\rangle>1.8 \, \hbar^2$ for
the lowest $35$ vibrational states of LiK.
They present a rich pendular dynamics, with significant contributions of
higher field-free rotational states. As $\nu$ gets closer to the dissociation
limit the relative weight of the $s$-wave becomes more important, and 
$\langle\mathbf{J}^2\rangle$ is reduced. In contrast, the field-free $J=0$
level still dominates the LiNa states with $\langle\mathbf{J}^2\rangle<0.26
\, \hbar^2$ for any $\nu$-value. 
Finally, let us remark that the effect of the field in terms of 
$\langle\mathbf{J}^2\rangle$ for the $\nu=95$ level of KRb is negligible 
but slightly increases again for the last states.

\section{Conclusion}
In conclusion, we have investigated the rovibrational spectra of the electronic
ground state $\textrm{X}^1\Sigma^+$ of the alkali dimers LiNa, LiK, LiRb,
LiCs, NaK, NaCs, KRb, and RbCs in the presence of a static homogeneous
electric field.  
Performing a comparative analysis, we
have considered the lowest rotational excitation within each vibrational band
up to the dissociation limit.
The vibrational state-dependent parameter $\omega_\nu$ provides a first
estimation of the influence of the field on the corresponding dimer  
and allows a classification of these systems into four groups. Thus, NaCs and
RbCs are most affected by the application of an electric field and LiNa is
least. 
In general, vibrationally highly excited states show less field impact
whereas already for moderate field strengths pronounced effects are 
observed for lower-lying levels. 
The richness and variety of field-dressed rotational dynamics have been
illustrated by analyzing the Stark shift of the spectra as well as the
orientation,  alignment, and hybridization of the angular motion. 

This work should serve as a guide to experimentalists interested
in electric field effects on cold and ultracold heteronuclear alkali dimers.
A natural extension of this work includes the study of the influence of
the field on states emerging from higher rotational excitations and/or with
nonzero magnetic quantum numbers $M$.

\acknowledgments
Financial support by the Spanish projects FIS2005--00973 (MEC) and 
FQM--0207, FQM--481 and P06-FQM-01735 (Junta de Andaluc\'{\i}a) is
gratefully appreciated.


\begin{thebibliography}{0}

\bibitem{speciss2004}
{Topical Issue on Ultracold Polar Molecules: Formation and
Collisions}, Eur. Phys. J. D {\bf 31}

\bibitem{kraft06} 
{Kraft S. D. {\it et al.}},
{J. Phys. B} {\bf 39}, {S993} (2006)

\bibitem{haimberger04} 
{Haimberger C. {\it et al.}},
{Phys. Rev. A} {\bf 70}, {021402(R)} (2004)

\bibitem{wang04} 
{Wang D. {\it et al.}},
{Eur. Phys. J. D} {\bf 31}, {165} (2004)

\bibitem{sage:203001} 
{Sage J. M. {\it et al.}},
{Phys. Rev. Lett.} {\bf 94}, {203001} (2005)

\bibitem{stann05} 
{Stan C. A. {\it et al.}},
{Phys. Rev. Lett.} {\bf 93}, {143001} (2004)

\bibitem{deh08} 
{Deh B. {\it et al.}},
{Phys. Rev. A} {\bf 77}, {010701(R)} (2008)

\bibitem{wille08} 
{Wille E. {\it et al.}},
{Phys. Rev. Lett} {\bf 100}, {053201} (2008)
        
\bibitem{hodby05} 
{Hodby  E. {\it et al.}},
{Phys. Rev. Lett} {\bf 94}, {120402} (2005)

\bibitem{ospelkaus} 
{Ospelkaus S. {\it et al.}},
{Phys. Rev. Lett} {\bf 97}, {120402} (2006)

\bibitem{bodo02}
{Bodo E., Gianturco F. A. \and Dalgarno A.},
{J. Phys. B} {\bf 35}, {2391} (2002)

\bibitem{demille02}
{DeMille D.},
{Phys. Rev. Lett.} {\bf 88}, {067901} (2002)

\bibitem{yelin06}
{Yelin S. F., Kirby K. \and C\^ot\'e R.},
{Phys. Rev. A} {\bf 74}, {050301(R)} (2006)


\bibitem{krems06}
{Krems R. V.},
{Phys. Rev. Lett.} {\bf 96}, {123202} (2006)

\bibitem{gonzalez07}
{Gonz\'alez-F\'erez R., Mayle M. \and Schmelcher P.},
{Europhys. Lett.} {\bf 78}, {53001} (2007)

\bibitem{gonzalez07_2}
{Gonz\'alez-F\'erez R., Weidem{\"u}ller M. \and Schmelcher P.},
{Phys. Rev. A.} {\bf 76}, {023402} (2007)

\bibitem{gonzalez04} 
{Gonz\'alez-F\'erez R. \and Schmelcher P.},
{Phys. Rev. A} {\bf 69}, {023402} (2004)

\bibitem{gonzalez06}
{Gonz\'alez-F\'erez R., Mayle M. \and Schmelcher P.},
{Chem. Phys.} {\bf 329}, {203} (2006)

\bibitem{mayle06}
{Mayle M., Gonz\'alez-F\'erez R. \and Schmelcher P.},
{Phys. Rev. A} {\bf 75}, {013421} (2007)

\bibitem{martin01} 
{Martin F. {\it et al.}},
{J. Chem. Phys.} {\bf 115}, {4118} (2001)

\bibitem{fellows91} 
{Fellows C. E.},
{J. Chem. Phys.} {\bf 94}, {5855} (1991)

\bibitem{staanum06} 
 {Staanum P. {\it et al.}},
{Phys. Rev. A} {\bf 75}, {042513} (2007)

\bibitem{russier00} 
{Russier-Antoine I. {\it et al.}},
{J. Physics B} {\bf 33}, {2753} (2000)

\bibitem{Docenko06} 
{Docenko O. {\it et al.}},
{J. Phys. B} {\bf 39}, {S929} (2006)

\bibitem{pashov07} 
{Pashov A. {\it et al.}},
{Phys. Rev. A} {\bf 76}, {022511} (2007)

\bibitem{fellows99} 
{Fellows C. E. {\it et al.}},
{J. Mol. Spec.} {\bf 197}, {19} (1999)

\bibitem{korek00} 
{Korek M. {\it et al.}} ,
{Chem. Phys.} {\bf 256}, {1} (2000)

\bibitem{Derevianko01} 
{Derevianko A., Babb J. F. \and Dalgarno A.},
{Phys. Rev. A} {\bf 63}, {052704} (2001)

\bibitem{aymar05} 
{Aymar M. \and Dulieu O.},
{J. Chem. Phys.} {\bf 122}, {204302} (2005)

\bibitem{koto05} 
{Kotochigova S. \and Tiesinga E.},
{J. Chem. Phys.} {\bf 123}, {174304} (2005)

\bibitem{byers70} 
{Byers Brown W. \and Whisnant D. M.},
{Chem. Phys. Lett.} {\bf 7}, {329} (1970)

\bibitem{meyenn} 
{von Meyenn K.},
{Z. Physik} {\bf 231}, {154} (1970)
\end{thebibliography}
\end{document}